\documentclass[a4paper,aip,jcp,floatfix,10pt,groupedaddress,preprint]{revtex4-1}

\usepackage{graphicx}
\usepackage[utf8]{inputenc}
\usepackage{tabularx}
\usepackage{epsfig}	
\usepackage[breaklinks=true]{hyperref}
\usepackage[usenames,dvipsnames]{xcolor}
\usepackage{textcomp}
\usepackage{ulem}
\usepackage{url}
\usepackage{epstopdf}
\usepackage{amsfonts}
\usepackage[paperwidth=210mm,paperheight=297mm,centering,hmargin=2cm,vmargin=2.5cm]{geometry}

\begin{document}

\renewcommand{\baselinestretch}{1.0}

\begin{center}
 {\large{\textbf{Interpretation of X-ray Absorption Spectroscopy in the Presence of Surface Hybridization}}}
\end{center}

\vspace{0.6cm}
\author{Katharina Diller$^*$}
\email[]{Author to whom correspondence should be addressed. Electronic mail: katharina.diller@tum.de,}
\email[]{current address: Institute of Physics, \'{E}cole polytechnique f\'{e}d\'{e}rale de Lausanne, CH-1015 Lausanne, Switzerland}
\affiliation{Department Chemie, Technische Universit{\"a}t M{\"u}nchen, \mbox{D-85747 Garching, Germany}}
\author{Reinhard J. Maurer}
\email[]{current address: Department of Chemistry, Yale University, New Haven CT 06520, USA}
\affiliation{Department Chemie, Technische Universit{\"a}t M{\"u}nchen, \mbox{D-85747 Garching, Germany}}
\author{Moritz M\"uller}
\email[]{current address: CIC nanoGUNE Consolider, Tolosa Hiribidea 76, E-20018 Donostia - San Sebasti\'{a}n, Spain}
\affiliation{Department Chemie, Technische Universit{\"a}t M{\"u}nchen, \mbox{D-85747 Garching, Germany}}
\author{Karsten Reuter}
\affiliation{Department Chemie, Technische Universit{\"a}t M{\"u}nchen, \mbox{D-85747 Garching, Germany}}


\begin{abstract}
X-ray absorption spectroscopy yields direct access to the electronic and geometric structure of hybrid inorganic-organic interfaces formed upon adsorption of complex molecules at metal surfaces. The unambiguous interpretation of corresponding spectra is challenged by the intrinsic geometric flexibility of the adsorbates and the chemical interactions with the interface. Density-functional theory (DFT) calculations of the extended adsorbate-substrate system are an established tool to guide peak assignment in X-ray photoelectron spectroscopy (XPS) of complex interfaces. We extend this to the simulation and interpretation of X-ray absorption spectroscopy (XAS) data in the context of functional organic molecules on metal surfaces using dispersion-corrected DFT calculations within the transition potential approach. On the example of X-ray absorption signatures for the prototypical case of 2H-porphine adsorbed on Ag(111) and Cu(111) substrates, we follow the two main effects of the molecule/surface interaction on XAS: (1) the substrate-induced chemical shift of the 1s core levels that dominates in physisorbed systems and (2) the hybridization-induced broadening and loss of distinct resonances that dominates in more chemisorbed systems.
\end{abstract}

\maketitle

\let\thefootnote\relax
\footnotetext{\dag~Electronic Supplementary Information (ESI) available}
\footnotetext{\ddag~current: Institute of Physics, \'{E}cole polytechnique f\'{e}d\'{e}rale de Lausanne, Switzerland, E-mail: katharina.diller@tum.de}
\footnotetext{$\|$~current: Department of Chemistry, Yale University, United States}
\footnotetext{$\P$~current: CIC nanoGUNE Consolider, Donostia - San Sebasti\'{a}n, Spain}

\section{Introduction}
A detailed understanding of the geometry and chemical structure, the electronic level alignment, and reactivity of organic adsorbates\cite{Gottfried2015, Tautz2007, Seki2001, Barlow2003, Zhu2004, Schultz2016} provides important information for the application of hybrid inorganic-organic systems (HIOS) in organic solar cells, organic light-emitting diodes (OLEDs) or other molecular electronics devices.\cite{Lu2007, Morgenstern2011,Dodabalapur1997, Hoppe2004, Witte2004} X-ray photoelectron spectroscopy (XPS) and near-edge X-ray absorption fine-structure (NEXAFS, often also referred to as XANES) spectroscopy represent some of the most popular techniques to study the electronic structure and (with polarization-dependent NEXAFS) the adsorption geometries~\cite{Haehner2006} of metal-adsorbed organic molecules. For XPS, and even more so for NEXAFS, the interpretation of the spectroscopic signatures is a challenge though, and disentangling effects of electronic structure and adsorption geometry of complex functional adsorbates purely from experiment can be close to impossible.

\begin{figure}
 \includegraphics[width=0.55\textwidth]{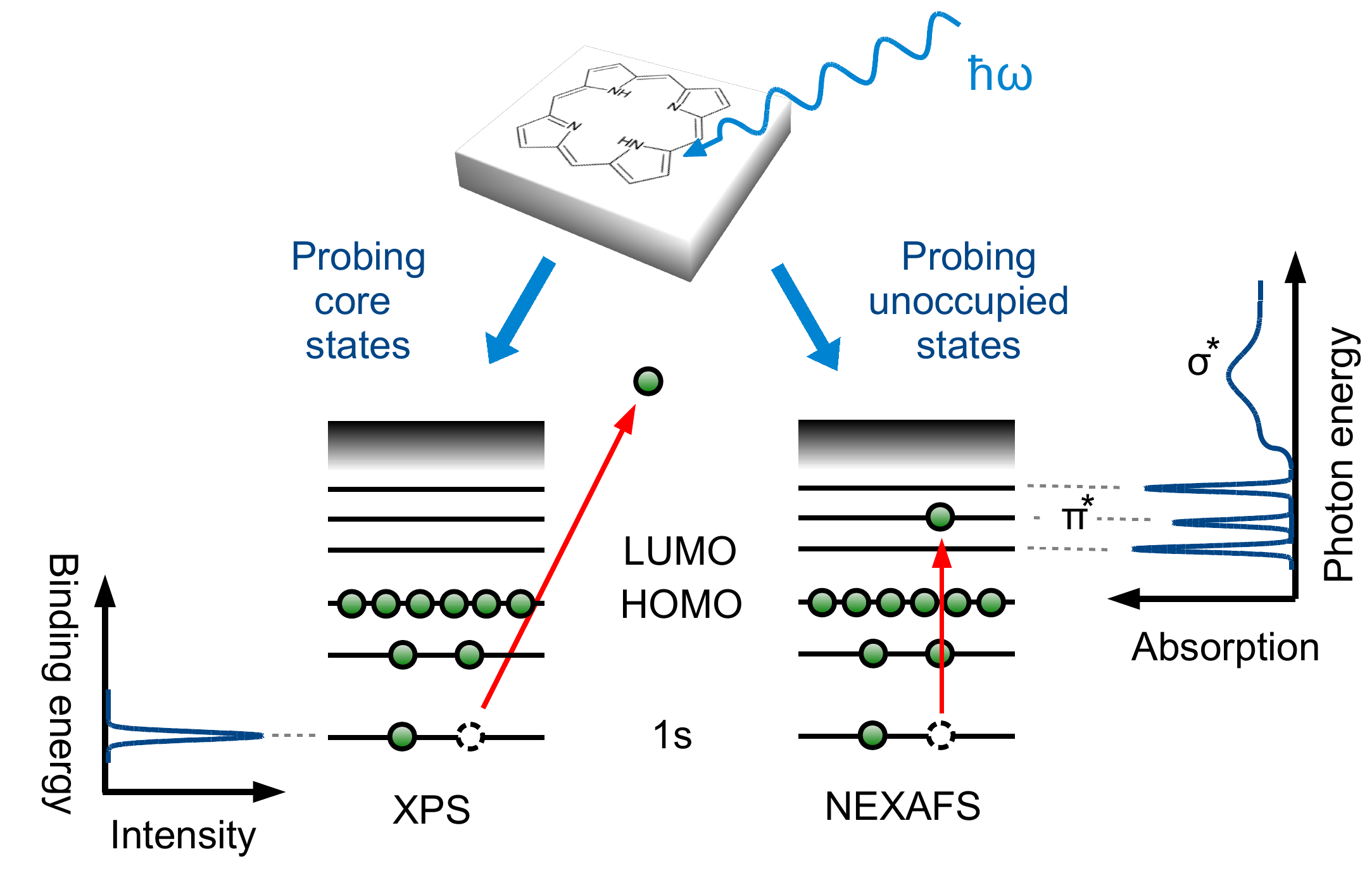}
  \caption{Principle of XP and NEXAFS spectroscopy. Absorption of X-ray radiation leads either to emission of core electrons (XPS) or to excitation into unoccupied molecular $\pi^*$ and $\sigma^*$ states (NEXAFS). The schematic spectra indicate the relation between adsorbate states and peaks in the spectra.}
  \label{fig:principle}
\end{figure}

The principles of XPS~\cite{Huefner2003} and NEXAFS~\cite{Stoehr1992} are depicted in Fig.\ \ref{fig:principle}: 
The sample is irradiated by an X-ray beam, leading either to the emission of photoelectrons from a core level (XPS)
or the excitation of core electrons to higher, unoccupied levels (NEXAFS). In both cases experimental spectra are acquired by scanning over an energy range: kinetic energies for XPS (which are linked via the photon energy to electron binding energies) and photon energies for NEXAFS (to probe resonant excitations). Ideally, the measured spectrum exhibits sharp signatures which can be directly assigned to originate from individual core levels (XPS, Fig.\ \ref{fig:principle} left) or from transitions between core levels and unoccupied states (NEXAFS, Fig.\ \ref{fig:principle} right). Correspondingly, changes in the chemical environment are reflected in (i) the position of the spectral features, as well as (ii) their shape and intensity. A full understanding of these signatures is therefore the key to understand the chemistry and the electronic properties of the combined molecule/surface system.

Unfortunately the inherent multi-peak structure in NEXAFS (Fig.\ \ref{fig:principle}b) makes interpretation difficult. Already for simple organic molecules containing only a single carbon atom (such as CO) or only a single distinguishable chemical carbon species (such as C$_2$H$_4$ or benzene) the carbon K-edges are non-trivial and interpretation is vastly aided by simulations.~\cite{Pettersson1998,Morin2004,Fronzoni2012,Nilsson2000} Organic adsorbates currently studied in the context of molecular nanotechnology are considerably more complex. Molecules such as pentacene derivatives and alkanethiols,\cite{Breuer2015} alkynes,\cite{Klappenberger2015} porphyrins,\cite{Auwaerter2015,Gottfried2015} or azobenzene derivatives~\cite{Willenbockel2015,Gahl2016} have a high molecular flexibility and their conformation on surfaces depends on many factors. Polarization-dependent NEXAFS is commonly used to derive adsorption geometries,\cite{Breuer2015,Haehner2006,Nefedov2013} but the results are only reliable when all peaks are correctly assigned to subgroups of the molecule, obtained e.g. by comparison to (polarization dependent or independent) simulations. Complex organic adsorbates contain many chemically inequivalent carbon atoms. The measured C K-edge is thus composed of several non-trivial contributions, which not only differ in shape and position, but may each react differently to the presence of the surface.

First-principles NEXAFS simulations of complex interfaces (including the substrate) may prove indispensable, but became computationally feasible only recently. In contrast, single-molecule calculations using time-dependent DFT, as well as the here employed transition potential (TP) approach~\cite{Triguero1998,Slater1972} are well established and generally reproduce experimental gas-phase and multilayer data very well.~\cite{Schmidt2010,Kolczewski2001,Luo2001} Relaxation effects in the TP approximation are taken into account by including half a core hole, and the excitations at different atomic centers allow the direct assignment of spectral features to individual atoms and transitions. This approach (which we will use in the following as described in the methods section) has been successfully employed in single-molecule simulation for reproducing the near-edge region of experimental K-edge data with high accuracy (for details see for example works on the C edge of anthracene,\cite{Klues2014} the F edge of perfluoropentacene,\cite{Klues2016} or the C\cite{Brumboiu2013} and the O edge\cite{Brumboiu2015} of phenyl-C$_{60}$-butyric acid methyl ester performed with the DFT cluster code StoBe\cite{StoBe}), which then allows a detailed fitting of the measured fine structure.\cite{Diller2012} However, such single-molecule calculations lack the perturbing effect of the substrate on the signatures of adsorbates in the monolayer, i.e. they neglect the effects that define the electronic and optical properties of the HIOS interface. 

In this work we assess the role of such molecule/substrate interaction in the interpretation of NEXAFS signatures using dispersion-corrected DFT in combination with the TP approach. Explicitly accounting for the extended metal surface (similar to the study of Baby et al.\cite{Baby2015} and in contrast to finite cluster calculations, as used for example in refs.\ \citenum{Balducci2015} and \citenum{Shariati2013}), we study the XPS and NEXAFS signatures of two prototypical showcases for HIOS, free-base porphine (\mbox{2H-P}, Fig.~\ref{fig:xps_nexafs}a) on Ag(111) and Cu(111). \mbox{2H-P} represents the most basic compound in the group of porphyrins; molecules which attract large interest due to their ubiquitous presence in biology and their large chemical flexibility and versatility.~\cite{Auwaerter2015,Gottfried2015} In a recent study\cite{Mueller2016} we found that the molecule-surface binding is dominated by dispersion interactions on both substrates. However the charge rearrangement and surface hybridization is more physisorption-like on Ag(111) and more chemisorption-like on Cu(111).~\cite{Mueller2016} This subtle difference for otherwise equivalent features makes these systems an ideal showcase for the study of surface-interaction effects on NEXAFS signatures. Our detailed analysis of first-principles calculated NEXAFS spectra shows that stronger hybridization on Cu(111) than on Ag(111) translates into a loss of clearly characterizable adsorbate signatures.

\vspace{0.5cm}
\section{Computational Details}
All DFT calculations have been performed using the pseudopotential plane wave code CASTEP 6.0.1 \cite{Clark2005} and employing standard library ultrasoft pseudopotentials (USPPs) \cite{Vanderbilt1990} for the geometry optimizations and the generation of the molecular orbital projeted density of states (MO-PDOS). Electronic exchange and correlation were treated with the semi-local PBE functional \cite{Perdew1996} and a plane wave cutoff of 450 eV (400 eV) was used for the calculation of the adsorbates (isolated molecule). As adsorbate structures are strongly governed by van-der-Waals interactions, the semi-local functional was augmented with the pairwise-additive dispersion correction scheme vdWsurf of Tkatchenko and co-workers.\cite{Tkatchenko2009,McNellis2009} All calculations were performed with (6 $\times$ 6) (111) four-layered surface slabs of Ag and Cu with PBE-optimized lattice constants of 4.14 \AA\ (bulk Ag) and 3.63 \AA\ (bulk Cu). The vacuum was chosen to exceed 20 \AA\ and Brillouin-zone sampling was done with a 2$\times$2$\times$1 (4$\times$4$\times$1) Monkhorst–Pack grid \cite{Monkhorst1976} for geometry optimizations (electronic structure calculations). All molecular degrees of freedom were fully relaxed until residual forces fell below 0.025 eV/\AA. The energetically most favored adsorption sites were on both substrates the bridge sites (with respect to the center of the molecule). The porphine molecules adsorb parallel to the surface with only minor deformations. The difference in maximum pyrrole tilt angle between porphines adsorbed onto Cu and Ag is less than 4$^\circ$. A more detailed description of the optimized geometries, as well as an analysis of the adsorption energies can be found in ref.\ \citenum{Mueller2016}

Core-Level spectroscopy simulations have been performed using the ELNES module in CASTEP \cite{Mizoguchi2009}, a self-written post-processing tool, and on-the-fly generated core-hole excited USPPs including a full (half) core hole for the simulation of the XPS (NEXAFS) spectra. XPS energies for each chemical species have been calculated as the difference in total energies between core-hole excited state and ground-state. The XPS intensity is simply given by stoichiometry, assuming ideal and species-independent generation of photoelectrons. The NEXAFS simulations are performed using the transition potential (TP) approximation,\cite{Triguero1998} where the occupation of the initial state orbital (here: C 1s) is set to 0.5 (see also refs.\ \citenum{Klues2014,Klues2016}). This allows to calculate all transition energies in a single DFT calculation by determining the difference between the eigenvalue of the state with n = 0.5 and those of the unoccupied orbitals. The number of simulations is therefore the same for the simulation of XPS and NEXAFS spectra. To facilitate comparison to the experiment, spectra were then broadened with Gaussian functions of varying width: Up to 5 eV over the first transition a broadening $\sigma$ of 0.2 eV was used, followed by a linear increase to 2.0 eV up to 15 eV over the first transition to account for the reduced lifetime of the $\sigma^*$ resonances which leads to increasing widths.\cite{Stoehr1992} The sum of the shifted and broadened contributions resulted in the spectra displayed in Fig. 2 and Fig. 3. The spectra in Figs. 2 and 3 are referenced with respect to the first feature of lowest-energy species (marked in black). Further details on the calculations can be found in the Supplemental Material.

\section{Results and Discussion}
\begin{figure}
 \includegraphics[width=0.55\textwidth]{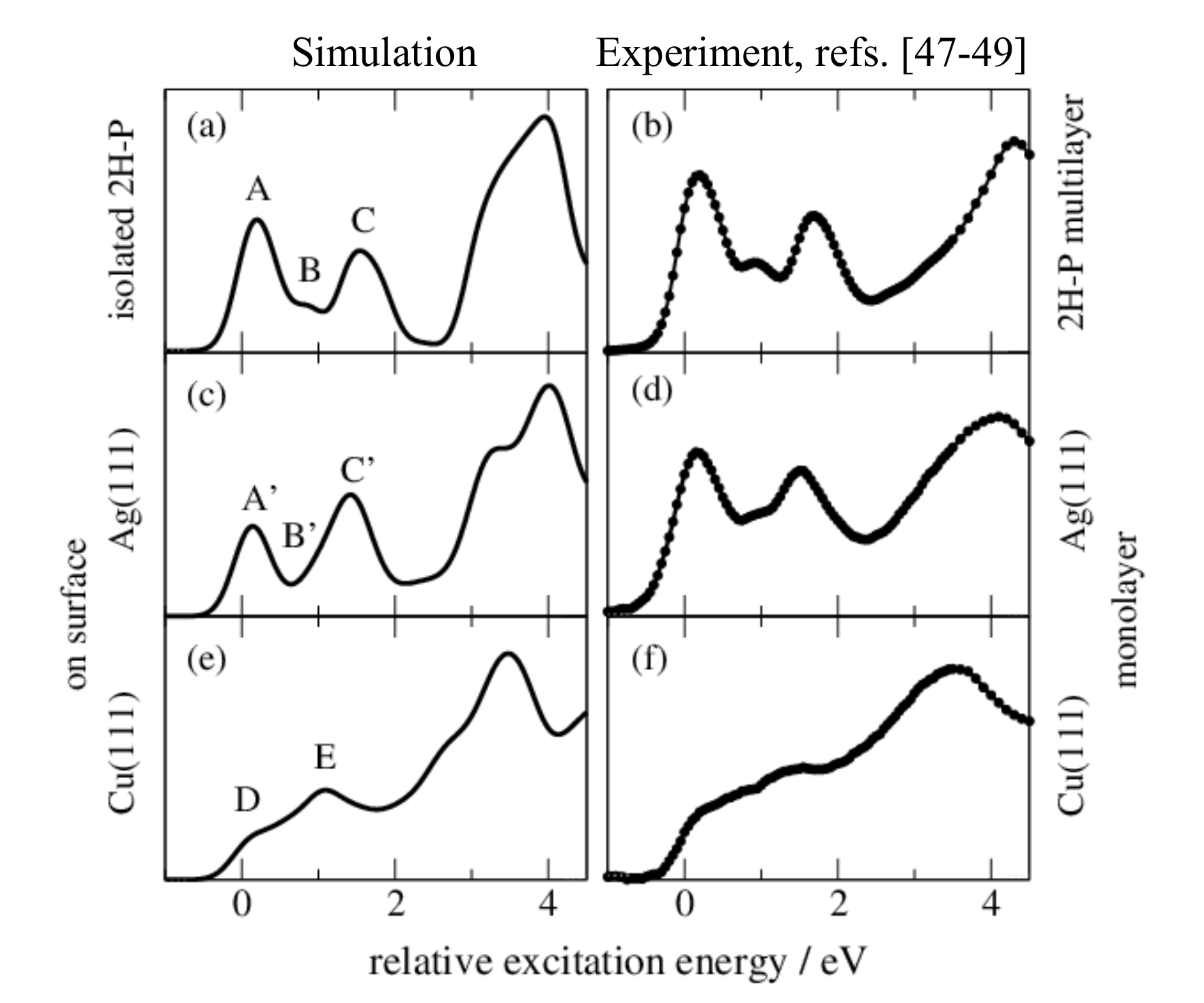}
  \caption{Comparison of simulated (left) and experimental (taken from refs.\ \citenum{Diller2014,Diller2013,Bischoff2013}) C K-edge NEXAFS spectra of \mbox{2H-P} (top), \mbox{2H-P} on Ag(111) (middle), and \mbox{2H-P} on Cu(111) (bottom).
  For ease of comparison both experiment and simulation have been shifted to align the leading edge with 0~eV, corresponding to an alignment to the lowest C-CN transition (cf.\ Fig.\ \ref{fig:xps_nexafs}). 
  }
  \label{fig:comp_exp}
\end{figure}

\begin{figure}
 \includegraphics[width=0.55\textwidth]{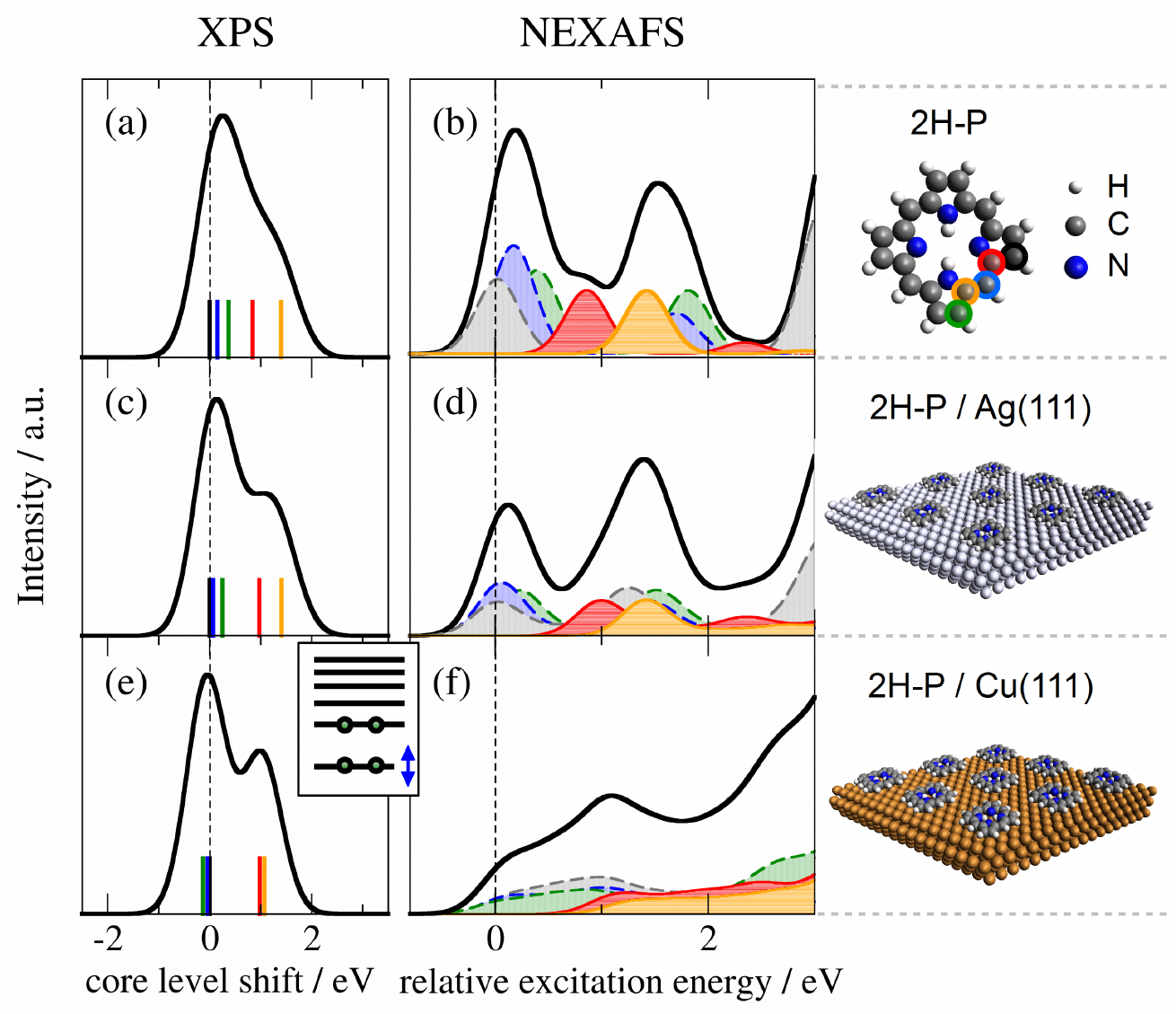}
  \caption{Comparison of simulated C1s XP (left) and C K-edge NEXAFS (middle) signatures of isolated \mbox{2H-P} (top) with those of \mbox{2H-P} adsorbed on Ag(111) (middle) and Cu(111) (bottom) surfaces. The color code of individual spectral components indicates the coordination of the corresponding C atom as indicated in the top right panel: C-bonded (black, blue, green dashed lines), N-bonded (red continuous line) and NH-bonded (orange continuous line). Total spectra (black lines) and NEXAFS components are broadened for visual clarity. For better comparison all spectra are aligned to the lowest C-CN transition (black component).}
  \label{fig:xps_nexafs}
\end{figure}

Figure~\ref{fig:comp_exp}b shows the experimental C K-edge NEXAFS multilayer spectra for \mbox{2H-P} on Ag(111).\cite{Bischoff2013} The spectrum is well reproduced by single-molecule calculations employing DFT cluster codes with the TP approach~\cite{Diller2014} and equally by our here employed pseudopotential plane-wave dispersion-corrected DFT calculations (Fig.\ \ref{fig:comp_exp}a) using a periodic supercell (cf.\ Methods and SI). The agreement between simulations and experiment allows for a direct assignment of the measured features in terms of transitions between core-levels and unoccupied molecular orbitals (MOs), cf. Fig.\ \ref{fig:xps_nexafs}b. Peak A is generated by transitions from the 1s levels to the $\pi^*$ lowest unoccupied MOs (LUMOs) in the presence of the respective core holes of the outer C-C bound carbon atoms. Peak B originates from the 1s $\rightarrow$ LUMO transition of the carbon atom attached to the iminic nitrogen atom, and peak C is a mix of several components (in agreement with previous single-molecule simulations\cite{Diller2014}).

\begin{figure}
 \includegraphics[width=0.5\textwidth]{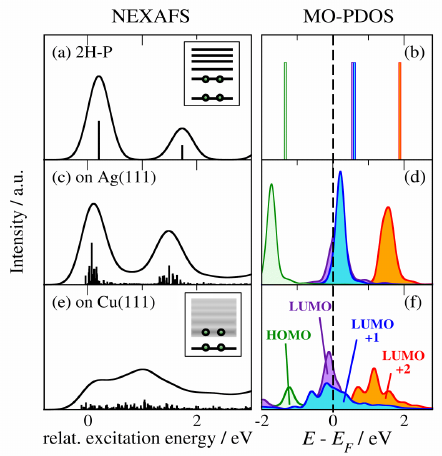}
  \caption{(Left) Simulated NEXAFS component originating from one C-bonded species (cf. blue dashed lines in Fig.\ \ref{fig:xps_nexafs}) for isolated 2H P (a), \mbox{2H-P} on Ag(111) (c), and on Cu(111) (e). Bars denote the calculated transitions. The continuous black line results after broadening. To increase visibility, bars in (c) and (e) are scaled by a factor of four compared to (a). (Right) The changes of the NEXAFS signatures reflect the differing hybridization in the molecular states as evidenced by the MOs projected from the total density of states (see text). Considering the discrete spectrum of isolated \mbox{2H-P}, no broadening was applied in (b).}
  \label{fig:pdos}
\end{figure}

For \mbox{2H-P} directly adsorbed on Ag(111) (Fig.\ \ref{fig:comp_exp}d, ref.\ \citenum{Bischoff2013}) the spectrum is very similar to that of the multilayer/isolated molecule. Only the second peak (peak B') seems shifted upwards leading to a broadening of the third (peak C'), and a reduction of the first peak (peak A'). In contrast, the strong modifications introduced by direct adsorption to Cu(111) (structures D, E in Fig.\ \ref{fig:comp_exp}e, refs.\ \citenum{Diller2013,Diller2014}) pose a challenge for any such interpretation. Just on the basis of the measured spectrum, it is not possible to unambiguously conclude whether all features are still present, but broadened; whether only the first peak is quenched due to an electron transfer to the LUMO \cite{Tseng2010}; whether the first peak is shifted upwards and the third one downwards; or even if the molecule is decomposed \cite{comment}. Using the detailed information on geometry, electronic structure, and individual atomic contributions provided by the first-principles calculations allows instead to clearly separate the spectral changes into two distinct effects: (i) chemical shifts of the 1s core levels (i.e., the XPS energies) which govern the onset of the NEXAFS curves of the different carbon species (cf.\ Fig.\ \ref{fig:xps_nexafs}), and (ii) the electronic structure of the frontier MOs, which determines the shape of the individual spectral components (cf.\ Fig.\ \ref{fig:pdos}). 

In the \mbox{2H-P} XPS signature (Fig.\ \ref{fig:xps_nexafs}a) the main low-energy feature is generated by purely carbon-coordinated C species, while the contributions from N-coordinated and NH-coordinated C species are clearly separated at higher binding energies. Upon adsorbing the molecule onto Ag(111) (Fig.\ \ref{fig:xps_nexafs}c) and Cu(111) (Fig.\ \ref{fig:xps_nexafs}e) the two groups separate more and more, up to the point of two displaced peaks in the case of \mbox{2H-P} on Cu(111) (for a comparison to experimental literature XPS data see Fig.\ S1). 
Accordingly, the first peak of the \mbox{2H-P} gas-phase NEXAFS curve originates from purely C-bonded carbon, the second peak from the C-N species, and the third is a mixture. The calculation confirms that the second peak (C-N component) is indeed shifted upwards upon adsorption on Ag(111) (Fig.\ \ref{fig:xps_nexafs}d). On Cu(111) (Fig.\ \ref{fig:xps_nexafs}f), however, the strong spectral changes can no longer be explained on the basis of chemical shifts alone.

This leads us to discuss the second adsorption-induced effect on the spectra that is not captured by single-molecule simulations: the variation of the shape of the individual spectral components arising from surface hybridization of the involved frontier MOs. To illustrate this, Fig.\ \ref{fig:pdos} compares the simulated NEXAFS spectra with the MO-projected partial density of states (MO-PDOS). For clarity we show only the component arising from one of the C-bonded species (blue dashed component in Fig.\ \ref{fig:xps_nexafs}), the others show the same behavior. In the MO-PDOS the frontier orbitals of isolated \mbox{2H-P} (Fig.\ \ref{fig:pdos}b) are projected onto the final eigenstates of the adsorbed system. The MO-PDOS thus exclusively reflects the changes of the frontier orbitals upon adsorption (see SI for more details). Figure \ref{fig:pdos}b shows the discrete frontier MOs of the isolated molecule. The corresponding NEXAFS spectrum in Fig.\ \ref{fig:pdos}a is governed by single transitions into the LUMO and LUMO+2. Due to the missing overlap between the 1s orbital and the LUMO+1 in the presence of the core-hole (cf. Supporting Information, Fig. S3) the corresponding transition has close to zero intensity. For \mbox{2H-P} on Ag(111) this discrete peak structure is mostly preserved, as well as the near-degeneracy of LUMO and LUMO+1. However, small hybridization effects already lead to broadened molecular states, and a partial shift below the Fermi level indicates an onset of charge transfer (which we recently confirmed using a range of charge partitioning schemes~\cite{Mueller2016}). This change in electronic structure carries over to the NEXAFS spectrum by replacing the delta-peak transitions of the isolated molecule with narrow distributions. The MO-PDOS alone, however, while reflecting the influence of the surface on the molecular states, is not sufficient for a detailed interpretation of the spectral changes. In Fig.\ S4 we compare the experimental data to the summed up MO-PDOS shown in Fig.\ \ref{fig:pdos}, which was multiplied by a step function to consider only the unoccupied states. The trends are well reproduced, but the near-edge fine-structure is not, which underlines the need for the inclusion of the core hole (in agreement with the results for single-molecule simulations presented for examples in refs.\ \citenum{Klues2014,Klues2016}) and a proper treatment of transition probabilities. The analysis of the experimental data does not allow an unambiguous decision whether the slight changes in spectral shape of the experimental NEXAFS data of \mbox{2H-P} adsorbed on Ag(111) compared to isolated \mbox{2H-P} are caused by a reduction of the first peak (caused by charge transfer) or a shift of the individual contributions.\cite{Bischoff2013} The here presented TP calculations show that the spectrum is not only governed by a relative peak shift, but also by a slight reduction of intensity in the first peak that stems from a charge transfer. The latter effect is evidenced by the MO-PDOS and the partial shift of the LUMO below the Fermi level (Fig.\ \ref{fig:pdos}d). In general, however, the quantification of transferred charge is not straightforward and results can differ depending on the computational setup.\cite{Hofmann2015} For a more detailed discussion of the amount of partial charge transfer between the Ag(111) and Cu(111) surfaces and the porphine molecule, the values obtained with different methods and the influence of the molecular coverage we refer to ref.\ \citenum{Mueller2016}.  

For the only modestly more reactive Cu(111) surface the MO-PDOS loses this discrete peak structure and instead exhibits band-like features due to hybridization with substrate states (Fig.\ \ref{fig:pdos}f). Rather than mere energetic shifts and significant broadening these features show a pronounced sub-structure with multiple peaks due to the splitting of MOs. These complex shapes again carry directly over to the NEXAFS spectrum (Fig.\ \ref{fig:pdos}e). The availability of a continuum of molecular states leads to a continuum of transitions with non-zero transition probability. In their superposition these transitions give rise to new features in the resulting spectrum that resemble peaks or shoulders. However, as is clear from the present analysis any association of these features to (predominant) transitions to specific resonant MOs would be unjustified, if not misleading: Neither can the spectral change be assigned to quenching of the first peak due to charge transfer alone, the molecule is neither decomposed nor deprotonated, nor is there a strong selective binding that would lead to a pronounced geometric deformation – as could all have been deduced by traditionally interpreting the spectrum in Fig.\ \ref{fig:xps_nexafs}f in terms of strong “shifts” or “quenching” of one or more of the three principal peaks of the gas-phase spectrum in Fig.\ \ref{fig:xps_nexafs}b. In fact, with the exception of the overall vertical adsorption height, the calculations yield only minimal differences in the optimized geometric structure of \mbox{2H-P} on Ag(111) and Cu(111), with the difference in maximum pyrrole tilt angle being less than 4$^\circ$.~\cite{Mueller2016} 

\vspace{0.5cm}
\section{Conclusions}
To summarize, we studied the influence of molecule-substrate interactions on X-ray absorption spectra of large organic adsorbates by using dispersion-corrected density-functional theory combined with the transition potential approach. For the example of free-base porphine adsorbed on the commonly employed substrates Ag(111) and Cu(111), this approach results in an excellent agreement between our X-ray absorption spectroscopy simulations and experimental literature data, with modest computational costs equivalent to commonly employed XPS simulations. By comparing the spectra of the adsorbed molecules to the spectra of the gas-phase reference (i.e., simulation of an isolated molecule), we could show that the implicit assumption of a small metal surface-induced perturbation of molecular states fails completely for the adsorption on copper. However also on the silver surface, typically considered as rather inert, assignment of all spectral changes necessitates simulation of the combined adsorbate/substrate system. For both HIOS we followed the two effects that shape the final C K-edge NEXAFS spectrum of the adsorbate: (1) Chemical shifts that are induced by the substrate-induced potential and (2) the broadening, splitting and complex shape of transitions into strongly hybridized frontier orbitals. The first effect can in principle be interpreted using a discrete MO picture, the second effect can not. While a surface-imposed broadening of molecular resonances is generally not surprising, its strength for a moderately reactive surface such as Cu(111) and an organic adsorbate that is predominantly bound by dispersive interactions is. The hybridization of the states modifies the NEXAFS spectrum up do the point where only a continuum of states is found. As illustrated by the showcase \mbox{2H-P} on Cu(111), what appear to be discernible spectroscopic signatures may then merely emerge from the superposition of a continuum of transitions with non-zero transition probability. This calls for utmost caution in the interpretation of adsorbate X-ray absorption spectra. Typical peak assignments in terms of adsorbate orbital symmetries ($\pi^*$,$\sigma^*$) or fitting procedures including only a few Gaussian or Lorentzian line shapes in such cases would not reflect the correct chemistry and could only be a rough approximation. Ideally, spectroscopic assignment should be supported by first-principles spectroscopic calculations that explicitly consider the effect of an extended surface. In lieu of such calculations, detailed XPS data can be equally helpful. Measured core-level shifts provide information about the energetic position of NEXAFS spectral components due to individual species. Even if nothing else is known about the hybridization and concomitant shape of the frontier orbitals, interpretation can then at least proceed in terms of atomic species.

\vspace{0.2cm}
\section*{Supplementary Material}
Additional computational information, a comparison to experimental XPS data, the total DOS, the final states orbitals, and a comparison between the experiment and the MO-PDOS are provided as Supporting Information.

\vspace{0.2cm}
\section*{Acknowledgments}
 The authors gratefully acknowledge computing time at the Leibniz Supercomputing Center (LRZ) under project pr85za. We thank F. Klappenberger, W. Auwärter, and J. V. Barth for helpful discussions. KD acknowledges support by the EPFL. RJM and KR acknowledge financial support by the Deutsche Forschungsgemeinschaft (DFG) under project \mbox{RE1509/16-2}.

\bibliography{references}

\begin{thebibliography}{52}
\bibitem [{\citenamefont {{Which does not seem likely for the example presented
  here, but could play a role for other complex compounds}}()}]{comment}%
  \BibitemOpen
  \bibfield  {author} {\bibinfo {author} {\bibnamefont {{Which does not seem
  likely for the example presented here, but could play a role for other
  complex compounds}}},\ }\href@noop {} {\ }\BibitemShut {NoStop}%
\end{thebibliography}
\bibliographystyle{aipnum4-1}

\end{document}